\documentclass[twocolumn,aps,prl,showpacs,amsmath,amssymb,psfrac,
superscriptaddress]{revtex4}
\usepackage{amsmath}

\usepackage{epsfig}
\usepackage{graphicx}
\usepackage{dcolumn}
\usepackage{bm}

\def\gl{\lower.35em\hbox{$\stackrel{\textstyle>}{\textstyle<}$}}
\def\gapp{\lower.35em\hbox{$\stackrel{\textstyle>}{\sim}$}}
\def\lapp{\lower.35em\hbox{$\stackrel{\textstyle<}{\sim}$}}

\begin{document}

\title{The ingap state spectral properties in the cuprates' pseudogap phase}
\author{Julius.\ Ranninger}
\affiliation{Institut N\'eel, CNRS et Universit\'e Joseph Fourier, BP 166, 
38042 Grenoble Cedex 09, France}
\author{Alfonso.\ Romano}
\affiliation{Laboratorio Regionale SuperMat, CNR-INFM, Baronissi (Salerno), Italy, and \\
Dipartimento di Fisica ``E. R. Caianiello'', Universit\`a di
Salerno, I-84081 Baronissi (Salerno), Italy}
\date{\today}

\begin{abstract}

The single-particle excitations, which initiate the pseudogap in the 
cuprate superconductors at some temperature $T^*$, relate to specific 
local spectral features of resonant pairing, where charge 
carriers get momentarily trapped on dynamically  deformable molecular 
Cu-O-Cu clusters. We show how those local excitations evolve into dispersive 
branches with a characteristic "S"-like shape for three-peak structured 
ingap excitations in the cuprates' pseudogap phase, when we consider a 
lattice of such clusters. This feature should be detectable in "momentum 
distribution curve" ARPES analysis.
\end{abstract}
\maketitle

{\it Introduction.} Superconductivity in the cuprates evolves out of an 
insulating state of phase uncorrelated pairs of charge carriers, when 
doping beyond a certain concentration of $x_{sc} = 5 - 10 \%$ holes  per 
Cu ion. Alternatively, for $x \geq x_   {sc}$, the superconducting state 
can be obtained out of the pseudogap state, which sets in below a certain 
$T^*(x)$. The opening of the pseudogap is due to phase uncorrelated 
fluctuating diamagnetic pairs. It is unrelated to any superconducting precursor pairing,\cite{Pasupathy-2008} which sets in  
below a certain temperature $T_M(x)\leq T^*(x)$ where these bonding pairs
acquire phase correlations on a finite space/time scale. This is evidenced 
in: a transient Meissner screening, the Nernst effect in thermal transport,
torque magnetization measurements and zero bias conductance 
of pseudogap/superconducting junctions.\cite{precursorpairing}
Our scenario of resonant pairing, attributed to the intrinsic
metastability of the cuprate crystal structure, involves                                                                                                                                                                                                                                                                                                                                                                                                   dynamically fluctuating molecular Cu-O-Cu 
clusters,\cite{Ranninger-2008,Ranninger-2010a, Ranninger-2010b} which, 
momentarily capture charge carriers  
in form of bound singlet pairs, respectively accommodate them  as itinerant particles
while passing through them. This causes a breaking of crystalline symmetry on a local level,\cite{Kohsaka-2008} which makes the fermionic charge carriers to be 
partly itinerant and partly localized, as  observed in STM imaging studies, see 
Ref.[\onlinecite{Lee-2009}] and references therein. A duplicitous feature of fermionic charge carriers,\cite{Hanaguri-2008} coexisting in localized and delocalized states,  had been  known for some time in dilute polaronic systems, in the region separating the anti-adiabatic and  adiabatic regimes \cite{Polaronworks}. It was this dual localization-delocalization feature, which lead one of us (JR) to conjecture in the early eighties, that in a Many Body polaronic system, it should result in an intrinsic  metastability of such compounds. It was a natural extension of the work on the Bipolaronic Superconductivity,\cite{Alexandrov-1981} an extremely fragile state of matter, most likely unrealizable in real materials, which prefer to localize their charge carriers under similar conditions.\cite{Chakraverty-1998}. In an attempt to materialize a superconducting state in strongly coupled electron lattice systems, it appeared judicious to consider the possibility of fluctuating bipolaronic states rather than bound states. This led  to  the proposition of a phenomenological model, the Boson Fermion model (BFM). A first attempt to explore this scenario \cite{Ranninger-1985} showed a pairing state below a certain temperature $T^*$, driven by a finite amplitude of the local pairing field. Such a state was later on ascribed to as the pseudogap state in the cuprates. Upon lowering the temperature this state can, but does not have to, condense into a phase correlated superfluid states. Depending on the concentration of charge carriers, it can equally result in an Mott correlation driven bipolaronic insulator.
A prime motivation for introducing this scenario was the widely appreciated fact that
metastable crystalline structures support fluctuating diamagnetic pairs, which  favored a superconductivity that avoided the stringent limitations of phonon mediated low temperature BCS Cooper pairing.\cite{Anderson-Matthias-1974} and thus could attain substantially higher critical temperatures $T_c$,\cite{Vandenberg-1977,Sleight-1991}.  The BFM, which captures the basic physics of polaronically driven pair fluctuating systems, can account for  the emergence of a phase correlated superconducting state of locally fluctuating bipolarons out of their correlation driven Mott-like insulating  
state.\cite{Cuoco-2006} The Boson-Fermion exchange mechanism proposed in this model avoids the condensation into the Bipolaronic Superconductor state. It preserves the fermionic character of the system, albeit with a fractionated Fermi surface \cite{Lee-2007,Kohsaka-2008} and  controls the breakdown of the superconducting state at $T_c$ by phase (i.e., via the disappearance of the density of superfuid carriers) rather than amplitude fluctuations. 

The pair fluctuation driven pseudogap, predicted on the basis of this model, \cite{Ranninger-1995} manifests itself 
in a wide temperature regime above $T_c$.  It exhibits ingap  states, whose nature is considered to play a  key role in our attempt to understand high $T_c$ superconductivity. The single-particle spectral 
function $A({\bf k},\omega)$ in this pseudogap phase in the temperature regime $[T_M,T^*]$, is predominantly determined by the strong {\underline local} phase correlations between itinerant and localized pairs of charge carriers on individual clusters, which make up the effective lattice sites of these cuprates. We show here how this local physics, which is at the origin of duplicitous localized and delocalized charge carriers, is carried over into the spatial (alias momentum) dependent $A({\bf k},\omega)$ and how the ingap states evolve into hybridized modes, constructed of (i) dispersionless branches, tracking the localized aspects and (ii) the bare tight binding spectrum of itinerant charge carriers, tracking their delocalized aspects.

{\it The scenario.} The BFM Hamiltonian, describing a lattice composed of effective sites which act as resonant trapping centers, is given by
\begin{eqnarray}
&& H  = \varepsilon_0 \sum_{i,\sigma}c^{\dagger}_{i\sigma}c_{i\sigma}
-t\sum_{\langle i j\rangle,\sigma}c^{\dagger}_{i\sigma}c_{j\sigma}
\nonumber \\
&& + \;  E_0 \sum_i \rho^{+}_i \rho^{-}_i
+ g \sum_i [\rho^{+}_i c_{i\downarrow} c_{i\uparrow}
+ c^{\dagger}_{i\uparrow} c^{\dagger}_{i\downarrow} \rho^{-}_i]. \quad
\label{Hamiltonian}
\end{eqnarray}
$c_{i\sigma}^{(\dagger)}$ denote  the annihilation (creation) operators
for fermions with spin $\sigma$ at some effective sites $i$ and
$\rho_i^{+}$ and $\rho_i^{-}$ are pseudo-spin 1/2 operators, describing tightly 
bound fermion pairs which behave as hardcore bosons. $t$, $g$, 
$\varepsilon_0 = zt - \mu$, $E_0 = \Delta_B -2\mu$  
denote respectively: the hopping 
integral for the fermions, the boson-fermion pair-exchange coupling constant, 
the local fermion and boson energy levels with respect to the chemical 
potential $\mu$, which has to be common to fermions and bosons, up to a factor 
2 for the bosons being composed of two fermions. The bosonic and fermionic 
particles deriving from the same source, requires that at any given moment, such 
locally fluctuating bosonic pairs and itinerant unpaired fermionic charge carriers 
are in thermal equilibrium. A transition from a phase correlated 
superconducting into the phase uncorrelated insulating phase takes place 
for a boson concentration close to $n_B = 1/2$, when increasing $g$ beyond a 
certain $g_{SIT}$. \cite{Cuoco-2006}
We shall here consider that case, taking $n_{tot} = n_F + 2 n_B = 2$ and  
study the cross-over at finite temperatures, from a metallic paramagnetic state into the diamagnetic pseudogap phase and eventually  into a correlation driven insulator of phase uncorrelated pairs by gradually increasing $g$.  
 
The prevailing physics of resonant pairing is related to the spectral properties of an  isolated site, which is encoded in the spectral properties of the atomic limit 
of eq. \ref{Hamiltonian}, $H_{at} = lim_{t \rightarrow 0} H$. 
The Hilbert space of this local problem consist of eight configurations of product states made out of four fermionic states $|2\rangle = |c^{\dagger}_{\uparrow} \rangle,\, |3\rangle = |c^{\dagger}_{\downarrow} \rangle, \,|6 \rangle =
|c^{\dagger}_{\uparrow} \rho^+\rangle, \, |7 \rangle = |c^{\dagger}_{\downarrow}\rho^+ \rangle$ with energies $E_2 = E_3 = \varepsilon_0,\, E_6 = E_7 = \varepsilon_0 + E_0$ and four bosonic states
$1 \rangle = | 0 \rangle,\, 
|4 \rangle \equiv |B\rangle = (1/\sqrt{2})[|c^{\dagger}_{\uparrow}c^{\dagger}_{\downarrow} \rangle  
- |\rho^+ \rangle], \,
|5 \rangle \equiv |AB \rangle = (1/\sqrt{2})[|c^{\dagger}_{\uparrow} c^{\dagger}_{\downarrow}\rangle 
+|\rho^+ \rangle], \,
|8 \rangle = |c^{\dagger}_{\uparrow}c^{\dagger}_{\downarrow}\rho^+ \rangle$ with energies $E_1 = 0,\, E_B = - g,\, E_{AB} = +g,\, E_8 =2 \varepsilon_0 + E_0$. In order to keep the algebra as simple as possible, we shall restrict 
ourselves to the discussion of the case, where the bosonic level coincides 
with  the center of the fermionic band i.e., $\Delta_B = 2\varepsilon_0$. This  
dictates the position of the chemical potential $\mu \simeq \Delta_B/2$ and 
implies a half-filled fermionic band ($n_F \simeq 1$). Putting
$\varepsilon_0 =0$, leads to $\Delta = \mu = 0$. 

The single-particle Green's function for this atomic limit has been derived previously, \cite{Domanski-1998,Domanski-2003c} and is:
 \begin{eqnarray}
 G_{at}(i\omega_n) &=& -\int_0^{\beta} d\tau \exp^{i \omega_n \tau}\langle
 T[c_{\sigma}(\tau)c^{\dagger}_{\sigma}]\rangle \nonumber \\ 
 &=& {Z^F \over i\omega_n - \varepsilon_0} + {1-Z^F \over i\omega_n - \varepsilon_0 - g^2/(i\omega_n + \varepsilon_0 - E_0)}, \quad 
 \end{eqnarray}
 with $Z^F = 2 /(3 + cosh \beta g)$. This local single-particle spectral 
 function $A_{at}(\omega) = -Im\, G_{at}(i\omega_n = \omega + i\delta)$ of an 
 isolated cluster site is composed of two contributions: 
 The first term arises from uncorrelated charge carriers. The second term 
 is reminiscent of the single-particle spectral function for BCS 
 superconductivity, with  $g$ playing the role of the gap. It describes the
 contributions coming from bonding  $|B\rangle$, respectively anti-bonding 
 states $|AB\rangle$. The 
 significant difference between resonant pairing  and BCS Cooper pairing  
 is the appearance in the spectral function of temperature dependent spectral 
 weights $Z_F$, respectively $1-Z_F$. It monitors the 
 relative weight of the intrinsic uncorrelated single-particle excitations 
 and those which derive from bonding and antibonding pairing states, as 
 we change $g$. We illustrate in Fig. \ref{A_at} the 
 variation with temperature of  $A_{at}(\omega)$.  As one 
 reduces the temperature  below a characteristic value $T \simeq g$, the 
 spectral intensity of the single-particle non-bonding excitations $Z^F$ at 
 $\omega = 0$,  shows a significant drop and goes to zero for
 $T\rightarrow 0$, while that of the bonding and anti-bonding states 
 at $\omega = \mp g$ increases correspondingly. Simultaneously the 
local exchange correlations, given by $\langle c_{\downarrow}c_{\uparrow} 
\rho^+ \rangle =(1/2)tanh(\beta g/2)$, show a marked increase upon decreasing $T$  
below $g$ and saturates at $0.5$ for $T \rightarrow 0$ (see Fig 3 in Ref. \onlinecite{Domanski-2003c}). These features 
indicate that the local density of states at the chemical potential ($\omega=0$) 
rapidly drops below $T=g$. Considering a finite system, it forshadows the opening 
of a gap-like structure at the Fermi surface, which is independent of any symmetry 
breaking. We shall now  explore, how this local physics, given by  
$A_{at}(\omega)$, evolves into dispersive modes, when  putting 
such effective sites into a lattice with charge exchange between adjacent sites. 
Such questions have to be handled in a non-perturbartive scheme, 
because of the strong inter-relation between single and two-particle excitations 
in resonant pairing systems. We choose for that purpose a dynamical mean field
theory (DMFT) analysis.
\begin{figure}[ht]
\begin{minipage}[c]{4.25cm}
\includegraphics[width=4.25cm]{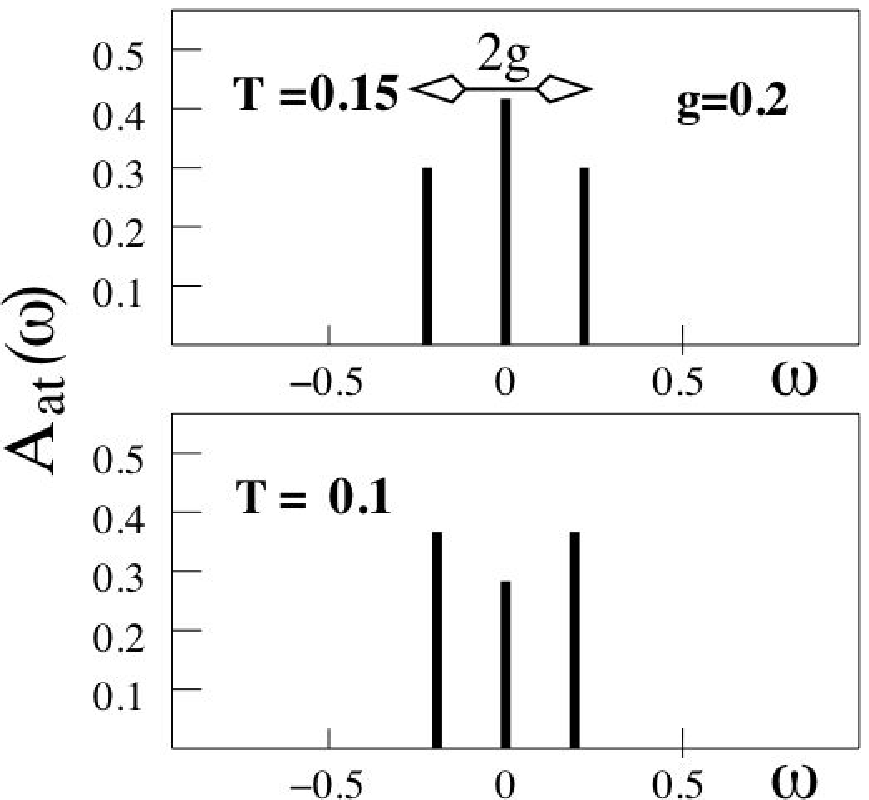}
\end{minipage}
\hfill
\begin{minipage}[c]{4.25cm}
\includegraphics[width=4.25cm]{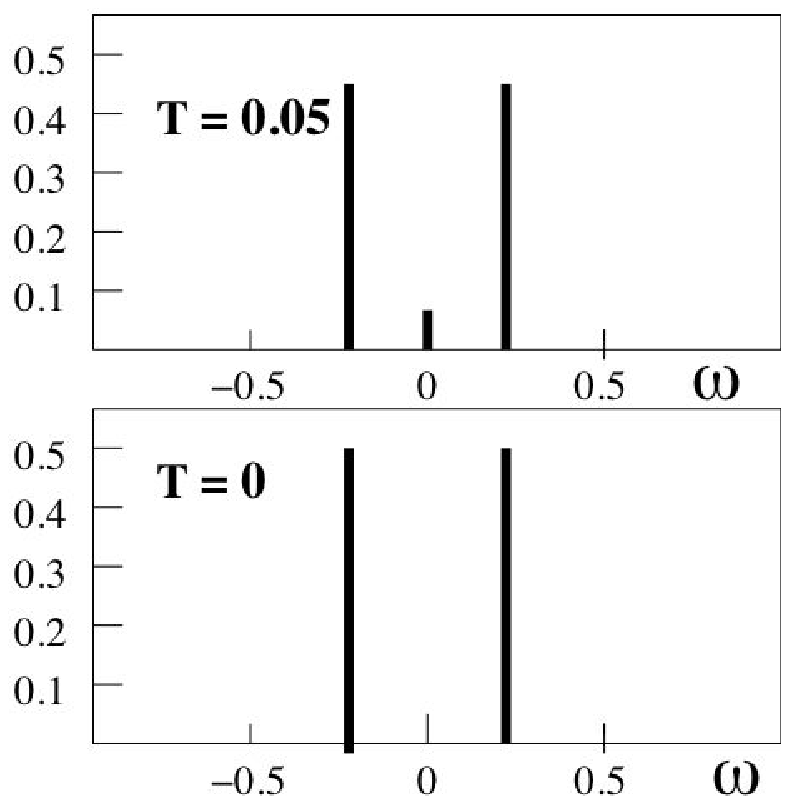}
\end{minipage}
\caption{The single-particle spectral function in the atomic limit  as a function of frequency for different temperatures $T$.}
\label{A_at}
 \end{figure}

 {\it Single-particle spectral features.}\,
 Taking into account the physics of the local problem ($H_{at}$) exactly, we 
 reformulate the Hamiltonian, Eq. 1, for the $D_{\infty}$ problem in the 
 standard way by coupling this local physics, contained in $H_{at}$ and 
 discussed above, to a Weiss field, which mimics the itinerancy of the 
 original fermions $c^{(\dagger)}_{i,\sigma}$ on a Bethe lattice. The 
 effective Hamiltonian for that is 
 \begin{eqnarray}
&H&\;=  \; g \; [ \; c_{\uparrow}^{\dagger}
c_{\downarrow}^{\dagger} \rho^- \; 
+ \; \rho^+ c_{\downarrow} c_{\uparrow}]  \nonumber \\
&+&\; \sum_{k,\sigma} w_{k} 
d_{k,\sigma}^{\dagger}d_{k,\sigma} \; + \; \sum_{k,\sigma} \; 
v_{k} \; [ \; d_{k,\sigma}^{\dagger}
c_{\sigma} \; + \; c_{\sigma}^{\dagger} d_{k,\sigma} \; ].
\end{eqnarray} 
$d_{k,\sigma}^{(\dagger)}$ denote the operators of the auxiliary 
Fermionic excitations of this Weiss field, having energies $w_{k}$ and which are 
coupled to the original fermionic excitations on an "Anderson impurity" site 
by a hybridization term of strength $v_k$. After rewriting this Hamiltonian 
in terms of Hubbard operators, corresponding to $H_{at}$ of the isolated local 
problem, we calculate the local Green's function
$G_{imp} (i\omega_n)=[i \omega_n - \varepsilon_0  - 
\Sigma_W(i\omega_n) - \tilde\Sigma^g_{int}(i\omega_n)]^{-1}$. Requiring 
$G_{imp} (i\omega_n)$ to be identical to the local part of lattice Green's function
$G_{lat}(i\omega_n)= \int d\varepsilon \,\rho(\varepsilon)
[i\omega_n - \varepsilon  - \widetilde{\Sigma}^g_{int}(i\omega_n)]^{-1}$, 
we obtain  $\Sigma_W(i\omega_n)=t^2 G_{imp}(i\omega_n)$, 
which determines the spectral distribution 
of the Weiss field energies $\{w_{k}\}$ in a selfconsistent way. 
$\rho(\varepsilon)=(1/2\pi t^2)\sqrt{\varepsilon(4t-\varepsilon)}$ denotes 
the density of states of bare uncoupled fermions of the lattice 
problem in $D_{\infty}$ with a band width $D=4t$. Once having 
obtained $\Sigma_W(i\omega_n)$, we deduce the selfenergy for the lattice
problem via $\tilde\Sigma^g_{int}(i\omega_n) = i\omega_n -\Sigma_W(i\omega_n)  - 
[G_{imp} (i\omega_n)]^{-1}$. In determining $G_{imp} (i\omega_n)$, we use 
a modification of the original Non-Crossing Approximation (NCA) approach by 
Bickers \cite{Bickers-1987}, adapted to the present BFM in
Ref.[\onlinecite{Romano-2000}]. By introducing
$\widetilde{\Sigma}^g_{int}(i\omega_n) \equiv 
\Sigma^g_{int}(i\omega_n)-\Sigma^{g=0}_{int}(i\omega_n)$ we  subtract 
out the effect of kinematic interactions, which arise in such a NCA 
formalism and are able in this way to describe qualitatively correctly the 
redistribution of spectral weight of the fermionic excitations over the 
entire frequency regime of the fermionic excitations.
\begin{figure}[ht]
\includegraphics[width=8cm]{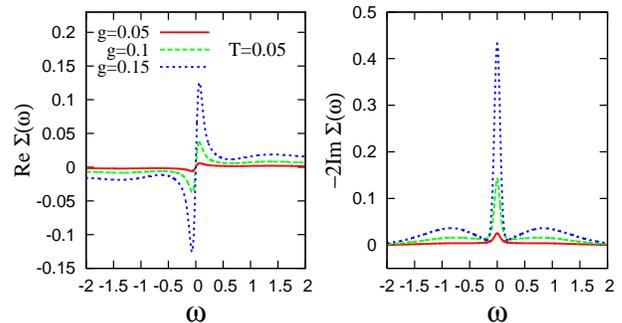} 
\caption{(Color Online) Real and imaginary part of  the fermionic self energy 
as a function of frequency $\omega$ for various values of $g$ and  $T= 0.05$ 
in units of $D$.}
\label{A_at}
 \end{figure}
We illustrate in Fig. 2 the real and imaginary part of the fermionic selfenergy  
for a given temperature $T=0.05$ and for three characteristic  values of $g$, which 
describe the (i) a paramagnetic metallic phase for $0.05 < g < 0.1$ and 
(ii) the pseudogap phase of diamagnetic bonding pairs, merging into a 
correlation driven insulator for $0.1 < g < 0.15$. The onset of the pseudogap
phase is manifest in the 
single-particle excitations, by the abrupt appearence above $g = 0.1$ of  a 
three-pole structure of the lattice Green's function, given by 
$i\omega  - \varepsilon - Re \widetilde{\Sigma}^g_{int}(i\omega) = 0$. 
This is illustrated in Fig. 3, where we plot the real ($\omega^*$) and imaginary 
($Im \Sigma(\omega^*$) parts of these poles. Upon approaching $g = 0.1$ from below, 
$\omega^*$ as a function of the bare fermionic spectrum $\varepsilon_{\bf k} =-tcos{\bf k}$ (presented here
by its corresponding energy $\varepsilon$, develops a kink-like structure around 
the chemical potential at $\varepsilon=0$ and finishes up in a vertical 
slope  upon approaching $g= 0.1$ $T=0.05$. With further increasing  $g$, we 
find  in a restricted region of momentum (alias $\varepsilon$) around  $\varepsilon=0$, simultaneously three dispersing modes. Two of them, the red continuous line and the blue dotted line outside the pseudogap, follow qualitatively the unrenormalized bare dispersion $\varepsilon_k \equiv \varepsilon$. Inside the pseudogap however, the renormalization turns the non-bonding states into a characteristic $S$-like shape (the green dashed line) in accordance with the three-pole structure of the atomic limit of the local spectral properties of $A(\omega)_{at}$. This $S$-like dispersing single-particle feature is a finger print of resonant pairing which unlike in the socalled cross-over scenario results not simply in bound pairs but in  strongly locally phase correlated localized  and delocalized states of bonding and anti-bonding states $|B\rangle$ and $|AB\rangle$.  Following the socalled Fermi-arcs in the Brillouin zone of the cuprate $CuO_2$ planar electronic structure, going from the nodal toward the anti-nodal point, whereupon the effective $g$ increases, the change-over from a well defined single ingap dispersive branch $\omega^*(\varepsilon) \simeq \varepsilon$ to an overdamped one dispersing in the $S$ like shape discussed above, should be clearly visible in a  
"momentum distribution curve" ARPES analysis. We plot it for that purpose  in Fig. 4  for a set of equidistant energies $\omega$, concentrating on the pseudogap energy region. Identifying the peak positions of the various curves, corresponding to different $\omega$'s, with the corresponding values of $\varepsilon$ we deduce the dispersion of those ingap excitations, which, as it should be, coincide with the dispersion of $\omega^*(\varepsilon)$ (see Fig. 3). 
\begin{figure}[ht]
\includegraphics[width=6cm]{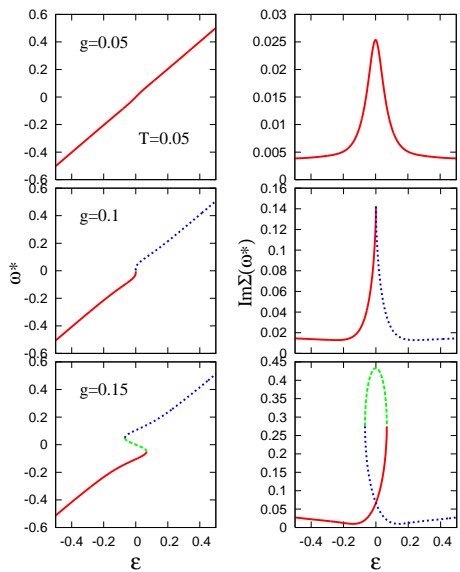} 
\caption{(Color Online) The real and imaginary parts of the poles of the 
single-particle Green's functions as a function of the bare energies 
$\varepsilon$ of the fermionic particles, measured from the chemical
potential and for a given temperature $T=0.05$ and different values of $g$
in units of $D$.}
\label{Fig3}
 \end{figure}
\begin{figure}[ht]
\includegraphics[width=8cm]{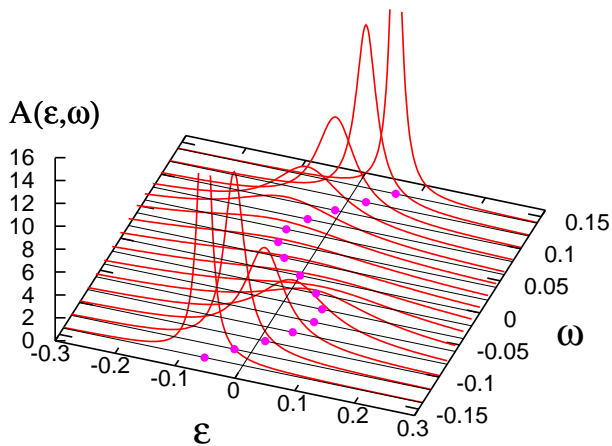} 
\caption{(Color Online) The single-particle spectral function 
$A(\varepsilon,\omega)$ as a function of the bare electron dispersion $\varepsilon$, scanning the various energies inside the pseudogap.}
\label{Fig4}
 \end{figure}

{\it Discussion} \, On the basis of a resonant pairing scenario and the Boson-Fermion model, where charge carriers are partly in localized pairing states and partly in delocalized single-particle states, we illustrated  how the local physics of such 
systems, derived from their intrinsic metastability, evolves into diffusive dispersing ingap states in the pseudogap phase of the cuprates. Such modes never appear as well defined  single-particle modes in such a resonant pairing scenario, in clear distinction to any BCS like physics \cite{Senthil-2009}. They are intrinsically  overdamped, with  a width extending over the entire pseudogap frequency region. Nevertheless, we can distinguish a characteristic $S$-like dispersion, which is imposed by the underlying local physics and which dictates the features of dispersing excitations. This is illustrated in presenting the single-particle spectral function  $A({\bf k},\omega)$ as a function of momentum ${\bf k}$ and  scanning the energies inside the pseudogap. Experimentally it should be possible to verify this by examining the "momentum distribution curves" of ARPES experiments and by going across the hidden Fermi surface near the antinodal points. 
In the present study we considered a temperature regime below the onset of the pseudogap at $T^*$, where fluctuating diamagnetic pairs on  molecular clusters are spontaneously created and destroyed.  These ingap states are best seen in this temperature regime, since below $T_M$ the onset of spatial superconducting phase correlations between the local diamagnetic fluctuations will strongly diminish  the spectral intensity of those ingap states, having shifted their spectral weight to emerging diffusive Bogoliubov branches.

{\it Acknowledgement} We thank Tadek Domanski for numerous discussions and comments on the manuscript.

\end{document}